\documentclass{article}
\usepackage{spconf,amsmath,graphicx}

\usepackage{graphicx}
\usepackage{amsmath}
\usepackage{amssymb}

\usepackage{epstopdf}
\usepackage{epsf}
\usepackage{fancyhdr}
\usepackage{hyperref}

\title{ An Accurate and Efficient Analysis of a MBSFN Network }
\name{ Salvatore Talarico and Matthew C. Valenti }
\address{West Virginia University, Morgantown, WV, USA.}

\begin{document}
\abovedisplayskip=0pt
\belowdisplayskip=0pt

\maketitle

\thispagestyle{empty}

\vspace{-0.3cm}
\begin{abstract}
\vspace{-0.1cm}
A new accurate analysis is presented for an OFDM-based multicast-broadcast single-frequency network (MBSFN). The topology of the network is modeled by a constrained random spatial model involving a fixed number of base stations placed over a finite area with a minimum separation. The analysis is driven by a new closed-form expression for the conditional outage probability at each location of the network, where the conditioning is with respect to the network realization. The analysis accounts for the diversity combining of signals transmitted by different base stations of a given MBSFN area, and also accounts for the interference caused by the base stations of other MBSFN areas. The analysis features a flexible channel model, accounting for path loss, Nakagami fading, and correlated shadowing. The analysis is used to investigate the influence of the minimum base-station separation and provides insight regarding the optimal size of the MBSFN areas.
In order to highlight the  percentage  of  the network that will fail to successfully receive the broadcast, the area below an outage threshold (ABOT) is here used and defined as the fraction of the network that provides an outage probability (averaged over the fading) that meets a threshold.
\end{abstract}

\vspace{-0.3cm}
\section{Introduction} \label{Section:Intro}
\vspace{-0.2cm}
Multicast-broadcast single-frequency network (MBSFN) is a transmission mode defined in the Long Term Evolution (LTE) standard \cite{ltea}, and in particular for the Multimedia Broadcast Multicast Service (MBMS) \cite{mbms2}. MBSFN is designed to send multicast or broadcast data as a multicell transmission over a synchronized single-frequency network (SFN).   A  group  of  those  cells  that are  targeted  to  receive  the same  data  is  called  an {\em  MBSFN area}. MBSFN enables the efficient delivery of applications such as mobile TV, radio broadcasting, file delivery and emergency alerts without the need of additional expensive licensed spectrum and without requiring new infrastructure and end-user devices. The transmissions from the different base stations in an MBSFN area are
tightly synchronized
and the MBSFN transmission appear to a user equipment (UE) as a transmission from a single large cell, with each base station transmission appearing as a separate  multipath component, dramatically increasing the signal-to-interference-noise ratio (SINR). Since only transmissions from base stations that lie outside the MBSFN area are interference, the intercell interference is reduced \cite{alexiou:2010}.

This  paper  presents  a  new  and  precise  analysis  for MBSFN orthogonal frequency-division  multiple  access  (OFDMA) networks. The analysis is driven by a new closed-form expression for the outage probability conditioned with respect to the network topology and shadowing. In particular, in \cite{torrieri:2012} a closed-form expression is derived for conventional networks, and in \cite{talarico:2013} it has been extended when signals arriving over different paths can be resolved in the presence of noise and interference. The  channel  model  accounts  for  path loss,  correlated shadowing,  and  Nakagami  fading,  and  the  Nakagami fading parameters do not need to be identical for all links.

In contrast with other works on MBSFN networks \cite{alexiou:2010,rong:2008,alexiou:2010b,alexiou:2012}, we don't use the classical approach to model cellular networks; i.e., we do not assume that the base stations are placed according to a  lattice  or  regular  grid.  Rather, we assume that the base-station locations are modeled as a realization of a random point process \cite{ganti:2009b,baccelli:2009,baccelli:2009b,andrews:2011,haenggi:2012}. In particular, in order to model the network more realistically, a {\em uniform cluster model} is used, which the authors have recently adopted to analyze both the downlink  \cite{Valenti:2013} and the uplink \cite{torrieri:2013} of a conventional non-cooperative cellular network. Having  adopted  a  realistic  model  and  an accurate  analysis, the influence of the minimum base-station separation is investigated and insight provided regarding the optimal size of the MBSFN areas. Furthermore, in  this paper we propose to quantify performance using the {\em area below an outage threshold} (ABOT),  which  is  here  defined  to  be  the fraction of the network that meets an outage constraint. The ABOT gives a useful  indication of  the  percentage  of  the network that will successfully receive the broadcast.


\vspace{-0.3cm}
\section{Network Model} \label{Section:SystemModel}
\vspace{-0.3cm}

The network comprises $M$ cellular base stations $\{X_1, ..., X_M\}$ placed on a finite square area $A_\mathsf{net}$ with sides of length $d_\mathsf{net}$. The variable $X_i$ represents both the $i^{th}$ base station and its location. To facilitate the analysis, the coverage area of the network is discretized into a large number of points, and the variable $Y_j$ is used to indicate the location of the $j^{th}$ point within the network.
 Each location $Y_j$ within a radio cell receives the same content from one or multiple base stations, which belongs to an MBSFN area. In each MBSFN area, the same content is synchronously broadcasted. Let $\mathcal G_{j,z}$ denote the set of the indexes of the base stations that belong to the $z^{th}$ MBSFN area and serving location $Y_j$. Furthermore, let $\mathcal Z_j$ denote the index of the MBSFN area that covers the location $Y_j$ and $N_j=|\mathcal G_{j, \mathcal Z_j}|$ be the number of base stations that belongs to that area.

The base stations are deployed according to a \emph{uniform-clustering} model \cite{torrieri:2012}. Using this model, the $M$ base stations are uniformly deployed in the network area $A_\mathsf{net}$.  An {\em exclusion zone} of radius $r_{bs}$ surrounds each base station, and no other base stations are allowed within this zone. In particular, each base station is placed one at a time uniformly into the portion of the network that remains outside of the exclusion zones of the previously placed base stations. The radius of the base-station exclusion zones can be primarily determined by economic considerations and the terrain.

The network comprises $S$ MFSFN areas, which are defined as follows. Inside the network arena, $\{Z_1, ..., Z_S\}$ points are picked according to a regular hexagonal grid and equally separated by $d_\mathsf{sfn}$, where the variable $Z_z$ represents the $z^{th}$ location. The $z^{th}$ MBSFN area is then formed by the radio cells of all base stations that are closer to the location of $Z_z$.

In a MBSFN OFDMA network, the transmissions are tightly synchronized and it is possible to combine the signals that are sent by all base stations $X_i, i \in \mathcal G_{j, \mathcal Z_j}$ and received at the UE, located at position $Y_j$, if the signals arrive within the  \emph{extended cyclic prefix}, which is fixed to be equal to $T_{ECP}=16.7 \mu s$ \cite{dahlman:2011}. The signal from base station $X_i, i \in \mathcal G_{\mathcal Z_j}$ to the UE at location $Y_j$ is then included in the MRC combined signal passed to the demodulator, if  $||X_i-Y_j||< d_\mathsf{max}$,
where $d_\mathsf{max}= 5$ km, otherwise it results in inter-symbol interference (ISI). The set $\mathcal G_{j,z}$ contains the indices of those base stations that are closer to $Z_{\mathcal Z_j}$ and they are located such that their signals arrive at $Y_j$ within the extended cyclic prefix. The set $\mathcal G_{j,z}$ is then selected such that $i \in \mathcal G_{j, \mathcal Z_j}$ if $||Z_{\mathcal Z_j}-X_i||<||Z_s-X_i||$, $\forall s  \neq \mathcal Z_j$ and also if $||X_i-Y_j||<d_\mathsf{max}$.

Let $\rho_{i,j}$ represent the instantaneous received power of $X_i$ at location $Y_j$, which depends on the path loss, shadowing, and fading. We assume that the path loss has a power-law dependence on distance. In particular, for a distance $d \geq d_{0}$, the path-loss function is expressed as the attenuation power law
\begin{eqnarray}
f\left(  d\right)  = \left(  \frac{d}{d_{0}}\right)  ^{-\alpha},
\label{eqn:pathloss}%
\end{eqnarray}
where $\alpha\geq2$ is the attenuation power-law exponent, and it is assumed
that $d_{0}$ is sufficiently large that the signals are in the far field.

Let $d_{i,j} = ||X_i-Y_j||$ be the distance between base station $X_i$ and location $Y_j$.  The instantaneous power of the signal received at the location $Y_j$ from the base station $X_i$ is
\begin{eqnarray}
   {\rho}_{i,j}
    =
   P_0 g_{i,j} 10^{\xi_{i,j}/10} f( d_{i,j} ), \label{eqn:instantaneous_power}
\end{eqnarray}
where  $P_0$ is the transmit power, which is common for all base stations, $g_{i,j}$ is the power gain due to fading and $\xi_{i,j}$ is a \textit{shadowing factor}.  The
$\{ g_{i,j} \}$ are independent with unit-mean, and $g_{i,j}=a_{i,j}^{2}$,
where $a_{i,j}$ is Nakagami with parameter $m_{i,j}$.
While the $\{g_{i,j}\}$ are independent from mobile to mobile, they are not
necessarily identically distributed, and for each mobile each link between $Y_j$ and $X_i$ can be characterized by a distinct Nakagami parameter $m_{i,j}$.
When the channel between $X_i$ and $Y_j$ experiences Rayleigh fading, $m_{i,j}=1$
and $g_{i,j}$ is exponentially distributed. It is assumed that the \{$g_{i,j}\}$ remain fixed for the duration of a MBSFN subframe, but vary independently from subframe to subframe.  In the presence of log-normal shadowing, the $\{\xi_{i,j}\}$ are zero-mean Gaussian with variance $\sigma_{s}^{2}$ and characterized by the normalized  autocorrelation  function  $\mathcal R\left( \Delta x\right)$,  where  $\Delta x$  is  the change in distance. The normalized  autocorrelation  function can be described with sufficient accuracy by an exponential function as \cite{Gudmundson:1991}
\begin{eqnarray}
\mathcal R\left( \Delta x\right)  =  \exp\left\{ -
\frac{||\Delta x||}{d_\mathsf{corr}} \ln 2 \right\}
\end{eqnarray}
with  the  decorrelation  length  $d_\mathsf{corr}$ ,  which  is  dependent  on  the
environment. For  the  urban  vehicular  test  environment (VTE),
\cite{mbms5}  proposes  $d_\mathsf{corr}=20$ m.  This  correlation  works  satisfactorily for  distances  up  to  approximately  $500$ m. In the absence of shadowing, $\xi_{i,j}=0$.

A \emph{distance-dependent
fading} model is assumed, where a signal originating at base station
$X_i$ arrives at location $Y_j$ with a Nakagami fading parameter $m_{i,j}$ determined as following
\begin{equation}
m_{i,j}=%
\begin{cases}
3 & \mbox{ if }\;||X_{i}-Y_{j}||\leq r_{\mathsf{f}}/2 \\
2 & \mbox{ if }\;r_{\mathsf{f}}/2<||X_{i}-Y_{j}||\leq r_{\mathsf{f}} \\
1 & \mbox{ if }\;||X_{i}-Y_{j}||>r_{\mathsf{f}}%
\end{cases}
\label{eqn:ddfading}
\end{equation}
where $r_{\mathsf{f}}$ is the \emph{line-of-sight radius}. The distance-dependent-fading model characterizes the situation where a mobile close to the base station is in the line-of-sight, while mobiles farther away are usually not.

\begin{figure*}[ht]
\belowdisplayskip=0pt
\vspace{-0.90 cm}
\setcounter{equation}{10}
\begin{eqnarray}
F_{\mathsf{Z}_j}(z \big| \boldsymbol \Omega_j)
 \hspace{-0.3 cm}& = &\hspace{-0.3 cm}
\sum_{k \in \mathcal G_{j,\mathcal Z_j} } \sum_{n=1}^{m_{k,j}} \Xi_{N_j} \hspace{-0.1 cm}\left( k, n, \left\{ m_{q,j} \right\}_{\forall q \in \mathcal G_{j,\mathcal Z_j}},\left\{\frac{\Omega_{q,j}}{\beta m_{q,j}} \right\}_{\forall q \in \mathcal G_{j,\mathcal Z_j}} \right) \left\{ 1 - \exp \left(\hspace{-0.1 cm}- \frac{\beta m_{k,j} z}{\Omega_{k,j}} \right) \sum_{\mu=0}^{n-1} \left(\frac{\beta m_{k,j} z }{\Omega_{k,j}} \right)^\mu \right. \vspace{-0.3 cm} \vspace{-0.4cm} \nonumber \\
\hspace{-0.3 cm} & & \hspace{-0.3 cm}
   \sum_{t=0}^\mu
\frac{z^{-t}}{\left( \mu - t \right)!} \hspace{-0.5 cm} \mathop{ \sum_{\ell_i \geq 0}}_{\sum_{i=0}^{M-N_j}\ell_i=t} \left.
\prod_{ i \notin \mathcal G_{j,\mathcal Z_j} }  \left[
 \ \frac{ \Gamma(\ell_i+m_{i,j}) }{ \ell_i! \Gamma( m_{i,j} ) }  \hspace{-0.1 cm}
\left( \frac{\Omega_{i,j}}{m_{i,j}} \right)^{\ell_i}
 \hspace{-0.1 cm} \left(
  \frac{\beta m_{k,j} }{\Omega_{k,j}} \frac{\Omega_{i,j}}{m_{i,j}} + 1
 \right)^{-(m_{i,j}+\ell_i)} \right] \right\}.
  \label{eqn_final_case1_Naka2}
\end{eqnarray}
\vspace{-0.15 cm}
{\hrulefill}
\end{figure*}
\begin{figure*}[ht]
\belowdisplayskip=0pt
\vspace{-0.65 cm}
\setcounter{equation}{11}
\begin{eqnarray}
 & &\hspace{-0.9 cm}\Xi_{L} \left( k, n, \left\{ r_q \right\}_{q=1}^{L},\left\{\eta_q \right\}_{q=1}^{L} \right) =
 \sum_{l_1=n}^{r_k} \sum_{l_2=n}^{l_1} \hspace{-0.1 cm} \cdots \hspace{-0.2 cm} \sum_{l_{L-2}=n}^{l_{L-3}}\hspace{-0.1 cm} \left[ \hspace{-0.1 cm} \frac{\left( -1 \right)^{R_L-r_k} \eta_k^n}{\prod_{h=1}^{L} \eta_h^{r_h}} \frac{\left( r_k+r_{1+u(1-k)}-l_1-1 \right)!}{\left( r_{1+u(1-k)} -1 \right)! \left( r_k - l_1\right)!}   \right. \nonumber \\
  & & \hspace{-0.9 cm} \left.
  \left( \frac{1}{\eta_j}-\frac{1}{\eta_{1+u(1-k)}} \right)^{l_1-r_k-r_{1+u(1-k)}} \frac{\left( l_{L-2}+r_{L-1+u(L-1-k)}-n-1 \right)!}{\left( r_{L-1+u(L-1-k)}-1\right)! \left(l_{L-2}-n\right)!}
  \left( \frac{1}{\eta_k} - \frac{1}{\eta_{L-1+u(L-1-k)}} \right)^{n-l_{L-2}-r_{L-1+u(L-1-k)}}  \right. \nonumber \\
  & & \hspace{-0.9 cm} \left.   \prod_{s=1}^{L-3} \frac{\left( l_s + r_{s+1+u(s+1-k)}-l_{s+1}-1\right)!}{\left( r_{s+1+u(s+1-k)}-1\right)! \left( l_s -l_{s+1}\right)!} \left(\frac{1}{\eta_k}-\frac{1}{\eta_{s+1+u(s+1-k)}}\right)^{l_{s+1}-l_s-r_{s+1+u(s+1-k)}}
\right]
   \label{Xi_L}.
\end{eqnarray}
\vspace{-0.10 cm}
{\hrulefill}
\vspace{-0.4 cm}
\end{figure*}
\setcounter{equation}{4}

The instantaneous SINR at location $Y_j$ by using (\ref{eqn:pathloss}) and (\ref{eqn:instantaneous_power}) can be expressed as \cite{rong:2008}
\vspace{-0.1cm}
\begin{eqnarray}
\gamma_j
&  = &
\frac{\displaystyle{\sum_{i \in \mathcal G_{j,\mathcal Z_j} } g_{i,j}\Omega_{i,j}}}
{\displaystyle\Gamma^{-1}
+
\sum_{ i \notin \mathcal G_{j,\mathcal Z_j} }
g_{i,j}\Omega_{i,j}},
\label{Equation:SINR2}
\end{eqnarray}
where $\Gamma=d_{0}^{\alpha} N_j P_0 /\mathcal{N}$ is the signal-to-noise ratio
(SNR) at a mobile located at unit distance when fading and
shadowing are absent, $\mathcal{N}$ is the noise power, and
\begin{eqnarray}
\Omega_{i,j}
& = &
\frac{10^{\xi_{i,j}/10} || X_i - Y_j||^{-\alpha} }{N_j }
\label{eqn:omega}
\end{eqnarray}
is the normalized power of $X_i$ at receiver $Y_j$.

\vspace{-0.4cm}
\section{Outage Probability} \label{Section:Outage}
\label{Section:OutageProbability}
\vspace{-0.3cm}
Let $\beta$ denote the minimum SINR required at location $Y_j$ for reliable reception and $\boldsymbol{\Omega }_j=\{\Omega_{1,j},...,\Omega _{M,j}\}$ represent the set of normalized base-station powers received at $Y_j$.  An \emph{outage} occurs when the SINR falls below $\beta$.  As discussed subsequently, there is a relationship between the SINR threshold and the supported {\em rate} of the transmission.  Conditioning on $\boldsymbol{\Omega }_j$, the outage probability of mobile $Y_j$ is
\begin{eqnarray}
   \epsilon_j
   & = &
   P \left[ \gamma_j \leq \beta \big| \boldsymbol \Omega_j \right].
   \label{Equation:Outage1}
\end{eqnarray}
Because it is conditioned on $\boldsymbol{\Omega }_j$, the outage probability depends on the particular network realization, which has dynamics over timescales that are much slower than the fading.
By defining
\begin{eqnarray}
  \mathsf S = \sum_{k \in \mathcal G_{j,\mathcal Z_j}} \beta^{-1} g_{k,j} \Omega_{k,j}, \hspace{0.2 cm} Y_i =g_{i,j} \Omega_{i,j}
\end{eqnarray}
\begin{eqnarray}
  \mathsf Z_j & = & \mathsf S  - \sum_{ i \notin \mathcal G_{j,\mathcal Z_j} }
  Y_i \label{eqn:z}
\end{eqnarray}
the conditional outage probability may be expressed as
\begin{eqnarray}
  \epsilon_j
  & = &
  P
  \left[
   \mathsf Z_j  \leq \Gamma^{-1} \big| \boldsymbol \Omega_j
  \right]
  = F_{\mathsf Z_j} \left( \Gamma^{-1} \big| \boldsymbol \Omega_j \right) \label{Equation:OutageCDF}
\end{eqnarray}
which is the cumulative distribution function (cdf) of $\mathsf Z_j$ conditioned on $\boldsymbol \Omega_j$ and evaluated at $\Gamma^{-1}$.  Restricting the Nakagami
parameter  $m_{i,j}$ between  location  $Y_j$ and  base station $X_i$ to be integer-valued, the cdf of $\mathsf Z_j$ conditioned on $\boldsymbol \Omega_j$ is proved in \cite{talarico:2013} to be (\ref{eqn_final_case1_Naka2}) at the top of this page. The function $\Xi_{L} \left( k, n, \left\{ r_q \right\}_{q=1}^{L},\left\{\eta_q \right\}_{q=1}^{L} \right)$ is defined by (\ref{Xi_L}) at the top of this page, where $u(x)$ is the step function and $R_L =  \sum_{k=1}^{L} r_k.$ \setcounter{equation}{12}

\vspace{-0.3cm}
\section{Network Performance}\label{Section:Results}
\vspace{-0.2cm}

\begin{figure}[t]
\centering
\includegraphics[width=8 cm]{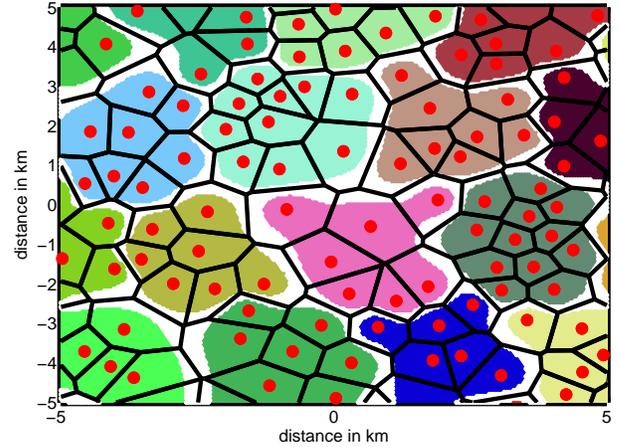}
\vspace{-0.4cm}
\caption{ Close-up  of  an  example  network  topology. The white areas are the portion of the network for which the outage probability is above a typical value of $\hat{\epsilon}=0.1$.
\label{Example} }
\vspace{-0.5cm}
\end{figure}

 In order to analyze the network performance, the {\em area below an outage threshold} (ABOT) is used, which is defined as the fraction of the network realization $t$ that provides an outage probability (averaged over the fading) that meets a threshold
\begin{eqnarray}
  \mathcal A_\mathsf{bot}^{\left(t\right)}
  & = &
 P \left[ \epsilon_j < \hat{\epsilon} \right].
  \label{area_below_an_outage_threshold}
\end{eqnarray}
The ABOT is an indicator of the percentage of the network that will successfully receive the broadcast. An outage threshold of $\hat{\epsilon} = 0.1$ is typical and appropriate for modern systems. For instance, in Fig.\ref{Example} it is shown a portion of an example network, where the area in white is the portion of the network for which the outage probability is above $\hat{\epsilon}=0.1$ with $\beta=0$ dB, $\Gamma=10$ dB and $\alpha=3.5$, which as expected corresponds to the edge of the MBSFN areas, that are here illustrated with different colors and obtained by fixing $d_\mathsf{sfn}=3$. The base-station  exclusion  radius  is fixed to $r_{bs} =  0.5$, $M=400$ base stations are deployed into a square network arena with side of length $d_\mathsf{net}=20$. The  base  station  locations  are given by the large filled circles and the Voronoi tessellation shows the radio cell boundaries that occur in the absence of shadowing.

After computing $ \mathcal A_\mathsf{bot}^{\left(t\right)}$ for $\Upsilon$ network topologies, its {\em spatial average} can be computed as following
\begin{eqnarray}
   \bar{\mathcal A}_\mathsf{bot}
  & = &
 \frac{1}{\Upsilon} \sum_{t=1}^{\Upsilon} \mathcal A_\mathsf{bot}^{\left(t\right)}.
  \label{area_below_an_outage_threshold}
\end{eqnarray}

A key consideration in the operation of the network is the manner that base stations select their rates at which the multimedia content is broadcasted. The SINR threshold depends on the modulation and coding scheme and receiver implementation. For a given $\beta$, there is a corresponding transmission rate $R$ that can be supported, and typically only a discrete set of $R$ can be supported. Let $R = C(\beta)$ represent the relationship between $R$, expressed in units of bits per channel use (bpcu), and $\beta$. While the exact dependence of $R$ on $\beta$ can be determined empirically through tests or simulation,  we make the simplifying assumption when computing our numerical results that $C(\beta) = \log_2(1+\beta)$ corresponding to the Shannon capacity for complex discrete-time AWGN channels. This assumption is fairly accurate for systems that use capacity-approaching codes and a large number of code rates and modulation schemes, such as modern cellular systems, which use turbo codes with a large number of available modulation and coding schemes.

\begin{figure}[t]
\centering
\includegraphics[width=8 cm]{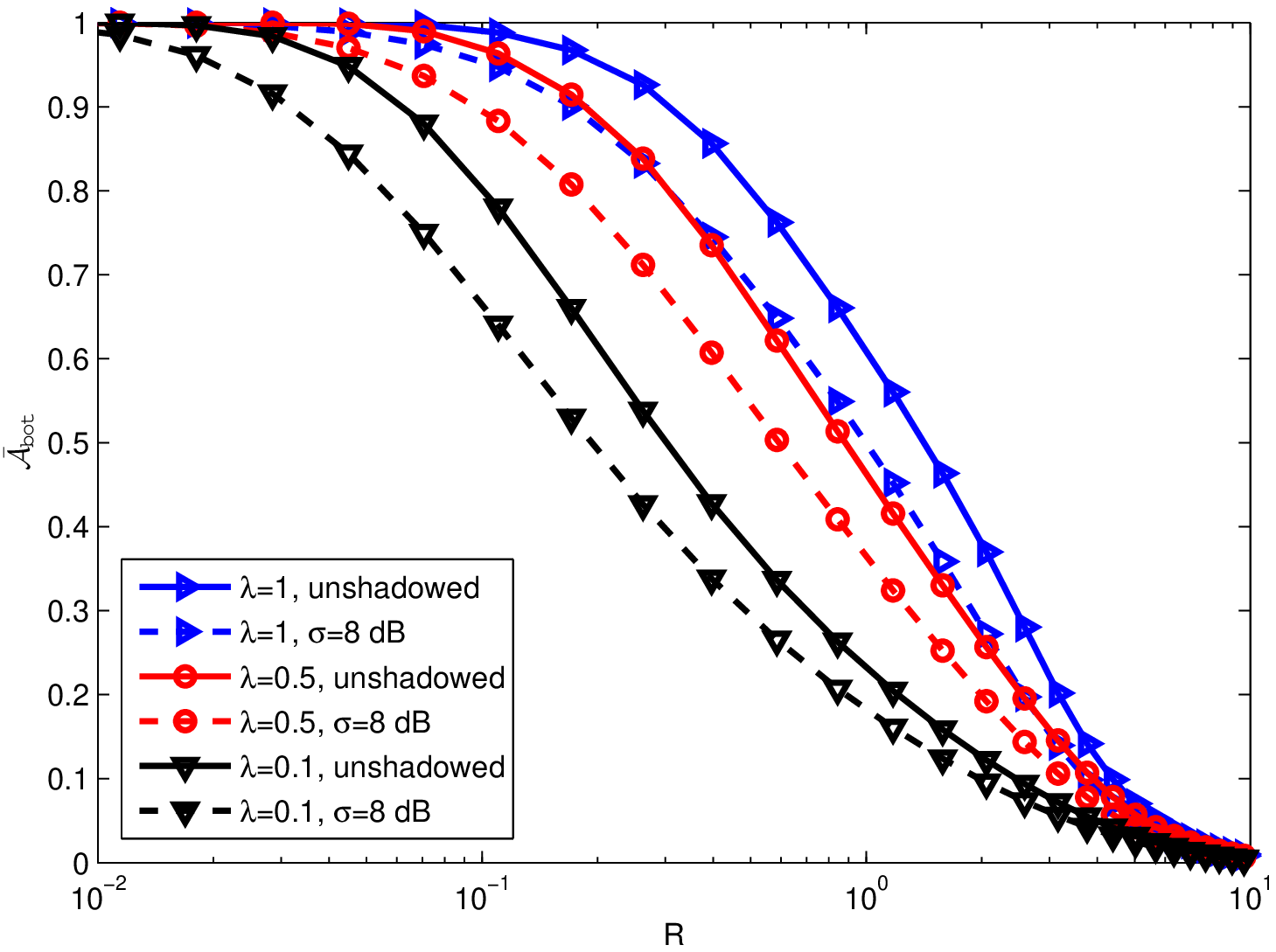}
\vspace{-0.4cm}
\caption{ ABOT as function of the rate for both a shadowed ($\sigma_s=8$ dB) and unshadowed environment. \label{ABOT_Rate_lambda}  }
\vspace{0.1cm}
\includegraphics[width=8 cm]{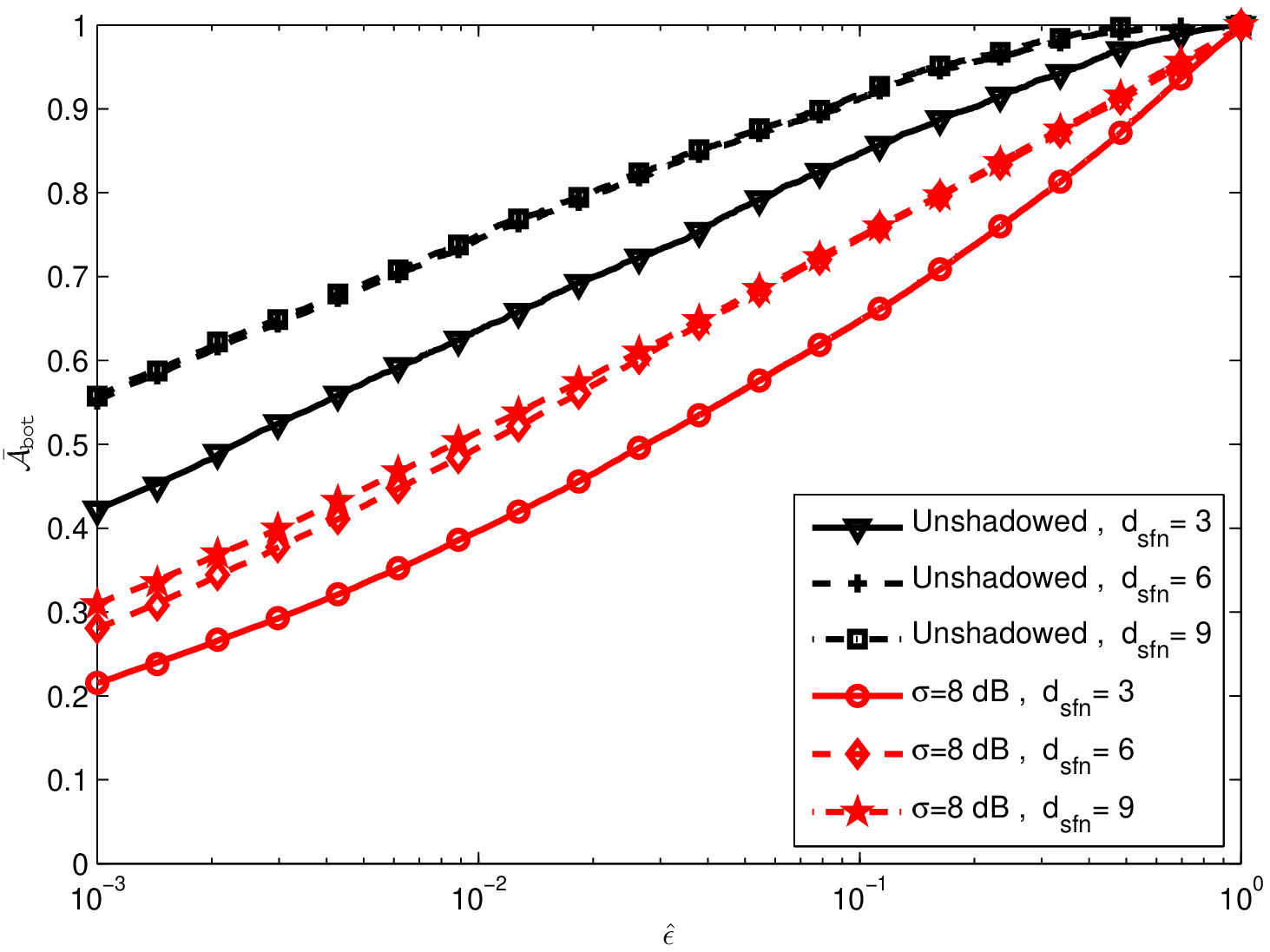}
\vspace{-0.5cm}
\caption{ ABOT  as a function of the threshold $\hat{\epsilon}$ for both a shadowed ($\sigma_s=8$ dB) and unshadowed environment with $R=0.5$.
\label{Abot_Shad} }
 \vspace{-0.55cm}
\end{figure}

\begin{figure}[t]
\centering
\includegraphics[width=8 cm]{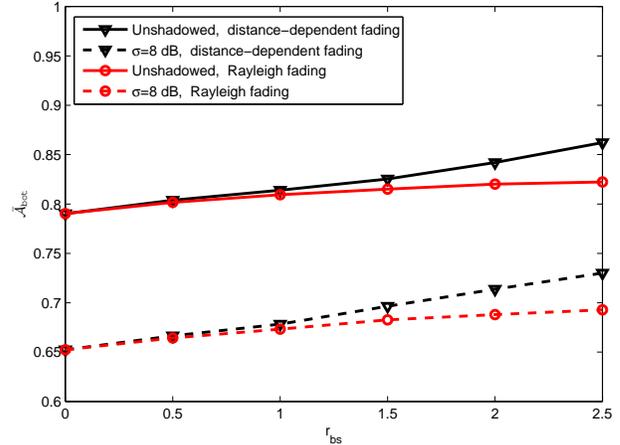}
\vspace{-0.4cm}
\caption{ ABOT as a function of $r_{bs}$ when $R=0.1$, $\lambda=0.1$ and $d_\mathsf{sfn}=6$. \label{Figure:A}  }
 \vspace{-0.5cm}
\end{figure}

In order to determine the rate $R$ for a typical network the following approach is used.
Draw a realization of the network by placing base stations according to a uniform clustering model with density $\lambda=M/A_{net}$. Group the base stations into MBSFN areas. Compute the path loss from each base station to a very large number of locations, selected such that they form an extremely dense grid that covers the entire network arena, applying randomly generated correlated shadowing factors if shadowing is present. Determine the outage probabilities by using (\ref{eqn_final_case1_Naka2}) parameterized by the SINR threshold for all locations. By applying the function $R=C(\beta)$, find the corresponding rate.

As an example, consider a square network area with sides of length $d_\mathsf{net}=20$, where the performance is evaluated only in the center $10 \times 10$ portion of the network, in order to exclude edge effects. The base-station exclusion zone is set to $r_\mathsf{bs} = 0.5$ and the outage constraint is set to $\hat{\epsilon} = 0.1$. The line-of-sight radius is $r_\mathsf{f} = 0.5$. Other  fixed  parameter  values  are $\alpha=3.5$, $\Gamma = 10$ dB and $\Upsilon=1000$.

Fig. \ref{ABOT_Rate_lambda} shows the ABOT as function of the rate for both a shadowed ($\sigma_s = 8$ dB) and an unshadowed scenario and for three values of lambda when $d_\mathsf{sfn}=6$: (1) $\lambda=1$, (2) $\lambda=0.5$ and (3) $\lambda=0.1$. Fig. \ref{ABOT_Rate_lambda} shows that there is a tradeoff between the two quantities. Shadowing is detrimental and an increase in the density of base-station results in an higher ABOT. Fig. \ref{Abot_Shad} shows the ABOT as function of the outage threshold  parameterized for three values of $d_\mathsf{sfn}$ when $\lambda=0.5$ and $R=0.5$.
Fig. \ref{Abot_Shad} shows that an increase in the size of the MBSFN areas results in an improvement in performance, but only until a certain value. After a certain value of $d_\mathsf{sfn}$, the ISI starts to increase and the regions at the edge of the MBSFN areas don't get any more benefit by increasing them furthermore.
Fig. \ref{Figure:A} shows the area below an outage threshold as function of the minimum separation among base-station when $R=0.1$,$\lambda=0.1$ and $d_\mathsf{sfn}=6$.
The curves are shown for both a shadowed ($\sigma_s = 8$ dB) and an unshadowed scenario and for both Rayleigh fading ($r_{f}=0$) and a distance-depending fading when $r_{bs}=r_{f}$. Fig. \ref{Figure:A} shows that the fraction of network that succeeds to meet an outage constraint increases when the base-stations have an higher minimum separation and this effect is more prominent under a distance-depending fading.

\vspace{-0.3cm}
\section{Conclusion} \label{Section:Conclusion}
\vspace{-0.2cm}
This paper has presented a new approach for modeling and
analyzing the performance of multicast-broadcast single-frequency network (MBSFN).
The  analysis combines a new outage probability expression, which is exact
for  a  given  network  realization,  with  a  constrained  random
spatial model, which allows the statistics to be determined for a
class of network topologies. The results show that an increase in the size of an MBSFN areas leads to an improvement in performance only until a certain value of $d_\mathsf{sfn}$ is reached and as expected performance increases as the minimum separation among base-station gets higher.

\newpage
\bibliographystyle{IEEEbib}
\bibliography{icassp2014}

\end{document}